\documentclass[reprint,prl,amsmath,amssymb,aps,superscriptaddress,floatfix]{revtex4-2}

\usepackage{natbib}
\usepackage{graphicx}
\usepackage{dcolumn}
\usepackage{bm}
\usepackage{braket}
\usepackage{ulem}
\usepackage{xcolor}

\DeclareSymbolFont{symbolsC}{U}{txsyc}{m}{n}
\DeclareMathSymbol{\Searrow}{\mathrel}{symbolsC}{117}

\definecolor{forestgreen}{rgb}{0.13, 0.55, 0.13} 

\begin{document}


\title{Robust and Efficient High-dimensional Quantum State Tomography}

\author{Markus Rambach}
\email{m.rambach@uq.edu.au}
\affiliation{Australian Research Council Centre of Excellence for Engineered Quantum Systems}
\affiliation{School of Mathematics and Physics, University of Queensland, QLD 4072, Australia}

\author{Mahdi Qaryan}
\affiliation{Australian Research Council Centre of Excellence for Engineered Quantum Systems}
\affiliation{School of Mathematics and Physics, University of Queensland, QLD 4072, Australia}

\author{Michael Kewming}
\affiliation{Australian Research Council Centre of Excellence for Engineered Quantum Systems}
\affiliation{School of Mathematics and Physics, University of Queensland, QLD 4072, Australia}

\author{Christopher Ferrie}
\affiliation{Centre for Quantum Software and Information, University of Technology Sydney, NSW 2007, Australia}

\author{Andrew G. White}
\affiliation{Australian Research Council Centre of Excellence for Engineered Quantum Systems}
\affiliation{School of Mathematics and Physics, University of Queensland, QLD 4072, Australia}

\author{Jacquiline Romero}
\email{m.romero@uq.edu.au}
\affiliation{Australian Research Council Centre of Excellence for Engineered Quantum Systems}
\affiliation{School of Mathematics and Physics, University of Queensland, QLD 4072, Australia}

\date{\today}

\begin{abstract}
The exponential growth in Hilbert space with increasing size of a quantum system means that accurately characterising the system becomes significantly harder with system dimension $d$. 
We show that self-guided tomography is a practical, efficient, and robust technique of measuring higher-dimensional quantum states. 
The achieved fidelities are over $99.9\%$ for qutrits ($d{=}3$) and ququints ($d{=}5$),  and $99.1\%$ for quvigints ($d{=}20$)---the highest values ever realised for qudit pure states.
We also show excellent performance for mixed states, achieving average fidelities of $96.5\%$ for qutrits.
We demonstrate robustness against experimental sources of noise, both statistical and environmental. 
The technique is applicable to any higher-dimensional system, from a collection of qubits through to individual qudits, and any physical realisation, be it photonic, superconducting, ionic, or spin.
\end{abstract}

\maketitle

Quantum systems are often naturally high-dimensional, and indeed in most quantum information architectures it is necessary to collapse this dimensionality to realise qubits.
However there are many situations where high dimensionality is advantageous. 
For practical applications like quantum communication, implementing \textit{qudits} rather than qubits brings about higher information capacity~\cite{Erhard2018a} and enhanced robustness against eavesdropping~\cite{Bechmann2000}.  
For quantum computation, qudits can lead to efficient distillation of resource states~\cite{Campbell2012} and simplified gates~\cite{Lanyon2009}.  
For quantum foundations, qudits afford more robust violations of Bell~\cite{Collins2002} and entropic inequalities~\cite{Kewming2020}.
Qudits have now been realised in a variety of physical architectures including photon shape~\cite{Mair2001,Langford2004}, trapped ions~\cite{Klimov2003,Low2020}, superconducting circuits~\cite{Neeley2009,Kiktenko2014}, and colour-centre spins~\cite{Soltamov2019}.

In order to benefit from qudits, we need to be able to physically implement, control, and measure them. 
Measuring the full quantum state via quantum state tomography is particularly challenging, as the parameter space for qudits grows as $d^{2n}{-}1$, where $d$ is the qudit dimension and $n$ is the number of qudits~\cite{Banaszek2000,Thew2002,Altepeter2005}.
Alternative methods for state tomography that require less measurements have been proposed and implemented, including compressed-sensing~\cite{Gross2010,Liu2012} and adaptive~\cite{Mahler2013,Qi2017,Granade2017} tomography, but these are still computationally expensive and sensitive to noise and experimental errors. 
A recent proposal, self-guided quantum state tomography~\cite{Ferrie2014} (henceforth referenced as self-guided tomography), offers the promise of high accuracy and precision with less measurements, high robustness, and no post-processing, but has been tested only in low dimension (two qubits, $d{=}4$~\cite{Chapman2016}), and is untested---particularly in terms of robustness and sensitivity---for systems of high dimensionality.

\begin{figure}[t]
\includegraphics[width=1\columnwidth]{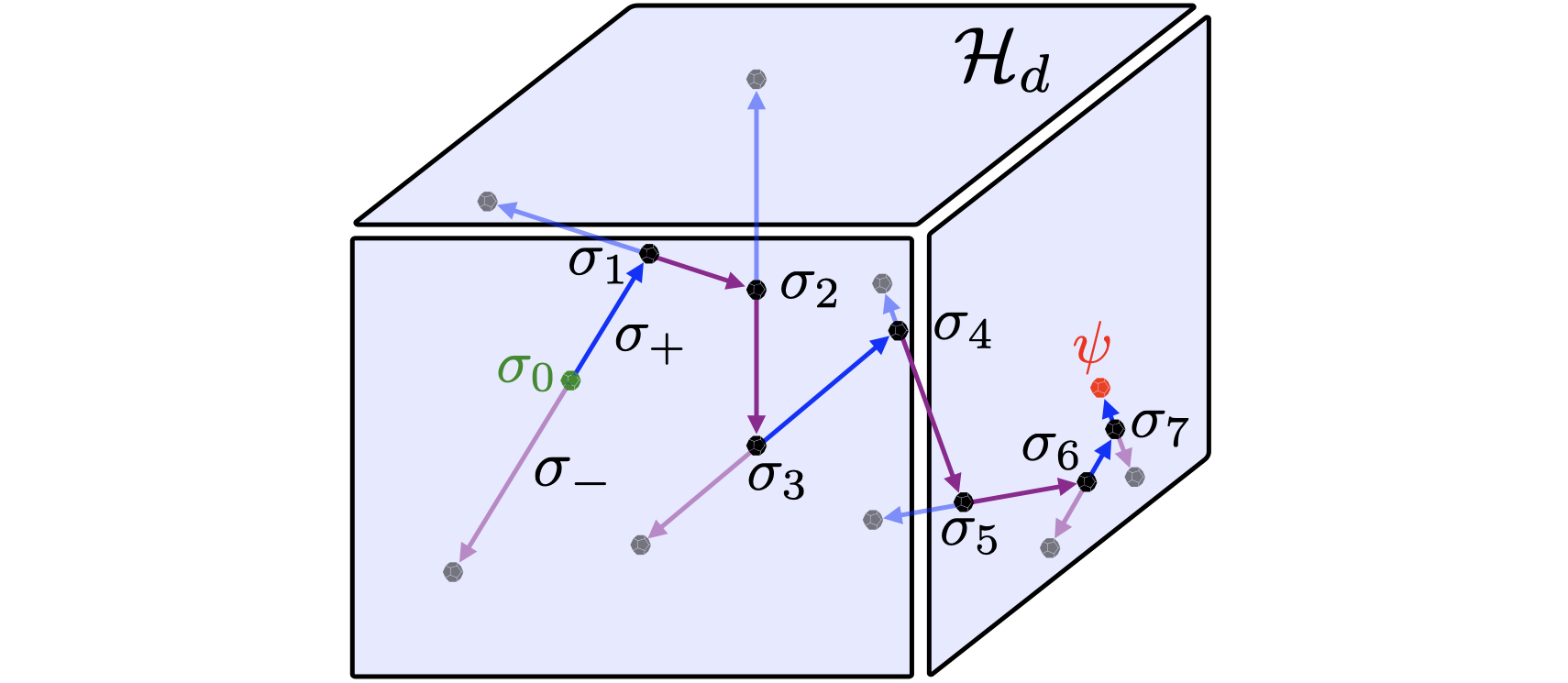}
\vspace{-6mm}
\caption{
\label{fig:path} 
Schematic representation of self-guided tomography for an arbitrary dimensional qudit in a Hilbert space $\mathcal{H}_d$. 
Starting from $\sigma_0$ (left, green), we are trying to estimate the unknown state $\psi$ (right, red). 
The algorithm calculates two directions, $\sigma_+$ in blue and $\sigma_-$ in purple, and determines their projections. 
The next state $\sigma_1$ is found by updating $\sigma_0$ according to Eq.~\ref{eq:sigma_k}. 
This process is repeated for a set number of iterations with decreasing step size, following the opaque path of highest overlaps, until a final state estimate (here very close to the unknown state $\psi$) is obtained. \vspace{-5mm} 
}
\end{figure}

Here we implement self-guided tomography on 3, 5, and 20 dimensional photon-shape qudits, achieving the highest measurement fidelities ever realised for qudits, from $99.92~\substack{+0.04 \\ -0.17}\,\%$ for qutrits, $d{=}3$, to $99.1\substack{+0.2 \\ -0.6}\,\%$ for quvigints, $d{=}20$.
We show that in high dimensions self-guided tomography outperforms standard tomography, both in terms of achievable fidelity for the same number of copies of the unknown state, and in overcoming errors due to mode-dependent losses, which often prevents standard tomography from converging on the true state. 
We test the robustness of self-guided tomography against statistical noise and atmospheric turbulence for different dimensionalities, and show excellent agreement between theory and experiment. 
Self-guided tomography will work with any qudit architecture: here we use photon-shape qudits because they are conveniently prepared and measured at room temperature.

In contrast to other algorithms, self-guided tomography solves the tomography problem by optimisation rather than estimation. 
It finds the true state by iteratively maximising the overlap between an unknown target state, $\ket{\psi}$, and a known estimate using only two measurements per iteration (see Fig.~\ref{fig:path}) \textit{independently} of the dimension of the system. 
This simple algorithm leads to a number of advantages like more economical scaling in the number of measurements, storage space, processing time, and robustness to various noise sources~\cite{Ferrie2014}.

In each iteration, the algorithm measures the estimation value $f(\ket{\sigma_\pm})$ of only two random states $\ket{\sigma_\pm}$ to approximate a gradient and update the estimated state. 
Fig.~\ref{fig:path} is a schematic of the algorithm's path for a Hilbert space $\mathcal{H}_d$, corresponding to an arbitrary qudit of dimension $d$. 
Starting from a current estimate $\ket{\sigma_k}$ self-guided tomography chooses one random direction~\cite{Utreras-Alarcon2019}, $(\Delta_k)_j \in \{1,-1,i,-i\}$, and measures the two projections $\ket{\sigma_\pm} {=} \ket{\sigma_k \pm \beta_k \Delta_k}$, shown in blue and purple. 

The system then calculates a gradient
\begin{eqnarray}
g_{k}{=}\frac{f\left(\sigma_{k}{+}\beta_{k} \Delta_{k}\right){-}f\left(\sigma_{k}{-}\beta_{k} \Delta_{k}\right)}{2 \beta_{k}} (\Delta_{k}^{-1})^\ast.
\label{eq:g_k}
\end{eqnarray}
Here, $\beta_{k}{=}b/(k{+}1)^t$ where $(b,t)$ are algorithm hyperparameters optimised via simulations.  
The value of $\beta_{k}$ decreases with the iteration number $k$ and is used to control the gradient estimation step size. 
The next estimate of the state is then updated following,
\begin{eqnarray}
\ket{\sigma_{k+1}}{=}\ket{\sigma_{k}{+}\alpha_{k} g_{k}}.
\label{eq:sigma_k}
\end{eqnarray}
Here, $\alpha_{k}{=}a/(k+1+A)^{s}$ where again $(a,A,s)$ are algorithm hyperparameters. 
The value of $\alpha_{k}$ controls the convergence and also decreases with $k$.
This procedure is repeated for each iteration by updating $\ket{\sigma_{k+1}} {\leftarrow} \ket{\sigma_k}$, resulting in a final state $\ket{\sigma_{f}}$, as illustrated in Fig.~\ref{fig:path} for $8$ iterations.

The values of the hyperparameters can depend on the dimension and specific system that is under investigation. 
However, we found the algorithm to be resilient against a variety of changes to their numerical values and chose similar values to Ref.~\cite{Ferrie2014,Chapman2016}; see the supplemental material S1~\cite{Supp} for results that highlight this resilience. 

At each step, the self-guided tomography is steered by the gradient $g_{k}$ and will thus converge to the underlying state after a sufficient number of iterations~\cite{Ferrie2014}. 
As our experimental results show, it is possible to arbitrarily increase the fidelity with the number of iterations.  
This can be chosen to match the required precision of the state estimation, which will generally depend on the application.

The quantum states in our experiment are encoded in the transverse shape of single photons, which we describe using the Laguerre-Gaussian basis $\{\ket{l_i,p_i}\}$~\cite{Allen2003}. 
Each mode in this basis is characterised by two numbers $\{l,p\}$, and the randomly chosen encoded state is given by the superposition $\ket{\psi} {=} \sum_i c_i \ket{l_i,p_i}$. 
In order to keep the spatial profile stable for each dimension $d$, i.e. the shape does not rotate upon propagation, we used modes that are of the same order by fixing the Guoy phase to $d{=}2p{+}|l|{+}1$.
We use a highly attenuated CW laser together with spatial light modulators and a single photon detector to implement a prepare-and-measure experiment, the schematic and details are given in S2~\cite{Supp}.

\begin{figure*}[t]
\includegraphics[width=\textwidth]{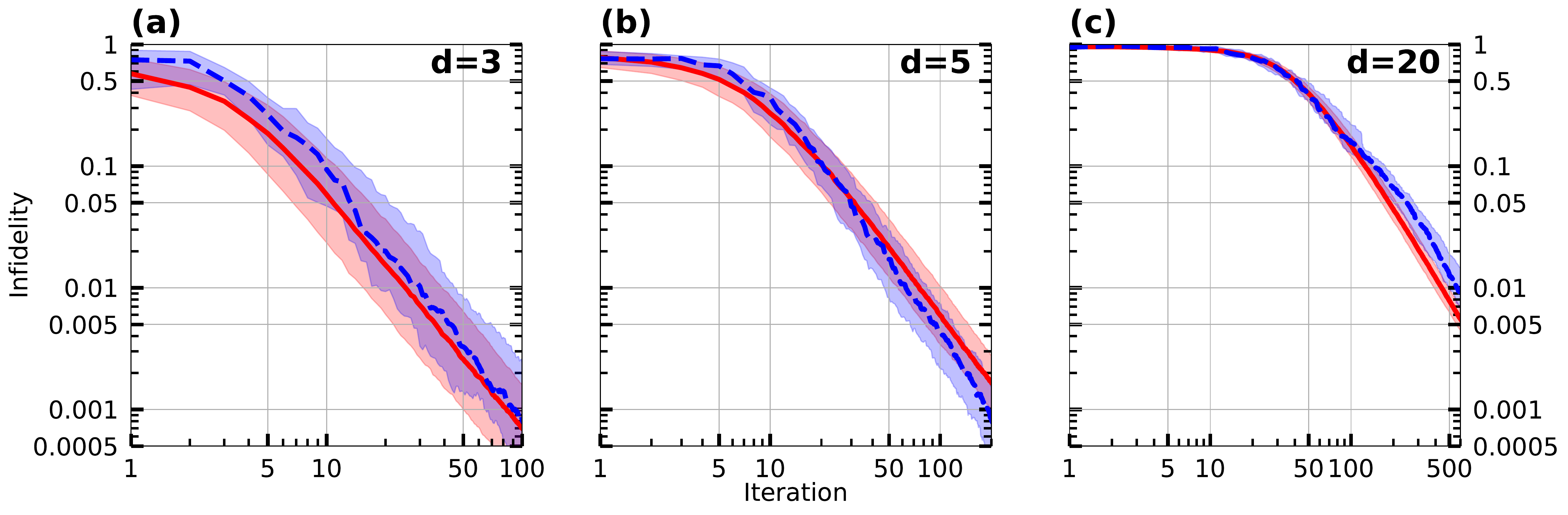}
\vspace{-6mm}
\caption{
\label{fig:SGQT_LN} 
Self-guided tomography for low statistical noise. 
Theoretical model in solid red and corresponding experimental result in dashed blue for (a) qutrit, (b) ququint, and (c) quvigint. 
The lines are the median performance, while the shaded regions are bound by the upper and lower quartile of the infidelities, $(50 \pm 25)\%$. 
The scaling of the horizontal axis is chosen to show where self-guided tomography surpasses infidelities below 0.1\% (a,b) and 1\% (c), respectively. The experimental results for $d{=}\{3,5\}$ are in high agreement with the theoretical predictions.
Note that the number of iterations required to achieve a given fidelity increases with the qudit dimension: for 99\% fidelity, 29, 62 and 566 iterations respectively. 
For quvigints, achieving this level of fidelity with standard over-complete tomography is not practically possible.
\vspace{-5mm}
}
\end{figure*}

The counts registered in the detector are proportional to the overlap $f(\sigma) {=} \left| \braket{\sigma|\psi} \right|^2$ and can be used directly in the algorithm by replacing parts of Eq.~\ref{eq:g_k}. 
That is:  
\begin{multline}
f\left(\sigma_+\right)-f\left(\sigma_-\right)=
f\left(\sigma_{k}+\beta_{k} \Delta_{k}\right)-f\left(\sigma_{k}-\beta_{k} \Delta_{k}\right)
\\
\Rightarrow\hspace{0.1cm}
\frac{\left| \braket{\sigma_{k}+\beta_{k} \Delta_{k}|\psi}\right|^2
- \left| \braket{\sigma_{k}-\beta_{k} \Delta_{k}|\psi}\right|^2}
{\left| \braket{\sigma_{k}+\beta_{k} \Delta_{k}|\psi}\right|^2
+ \left|\braket{\sigma_{k}-\beta_{k} \Delta_{k}|\psi}\right|^2}
= \frac{N_+ - N_-}{N_+ + N_-}
\label{eq:exp_SGQT},
\end{multline}

\noindent where $N_{\pm}$ are the number of counts from the two measurement states $\ket{\sigma_\pm}$ corresponding to measurement directions $\pm \beta_{k} \Delta_{k}$ for each iteration of self-guided tomography. 
Note that the denominator is taken as the sum of these counts. 
Given that $\ket{\sigma_\pm}$ are not orthogonal, this sum changes in each iteration.  
However, this pseudo-normalisation works because we are not interested in the actual values of the probabilities but rather the direction in which we  get more counts.

We investigated the performance of self-guided tomography for  qudits of dimensions $d{=}\{3,5,20\}$. 
We tested in three different regimes: low statistical noise, significant statistical noise, and weak turbulence.
In each case we compare the performance with standard quantum tomography using both root-approach and maximum-likelihood estimators~\cite{Bogdanov2009}, where we have constrained the standard tomography algorithm to return only pure states, thus ensuring a fair comparison with self-guided tomography. 
(Note, this constraint means that the figures returned from standard tomography will be the upper limit). 
This is applicable for systems---such as photons---where there is high degree of control in the preparation.   
For the true state $\ket{\psi}$, we use the estimate found from self-guided tomography after a much larger number of iterations. This is justified by the fact that our preparation fidelity---albeit high---is not 100$\%$.
More details can be found in S3~\cite{Supp}.

Fig.~\ref{fig:SGQT_LN} shows the results of self-guided tomography in the presence of low statistical noise, i.e. for high count rates. 
Fig.~\ref{fig:SGQT_LN}a and Fig.~\ref{fig:SGQT_LN}b show the results for qutrit ($\text{d}{=}3$) and ququint ($\text{d}{=}5$), respectively. 
Since our fidelities, $f(\sigma)$, are very high we plot the infidelity, $1 {-} \left| \braket{\sigma|\psi} \right|^2 {=} 1 {-} f(\sigma)$ for clarity. 
The experimental curve---the blue dashed line in Fig.~\ref{fig:SGQT_LN}a (Fig.~\ref{fig:SGQT_LN}b)---is derived from the median of 50 randomly chosen states of qutrits (ququints). 
To compare against our experiments, we simulate self-guided tomography on 1000 random qutrits (ququints) with the same count rate of ${\sim} 10^5$~Hz, dark count rate, and technical noise (e.g. due to alignment imperfections) as in our experiment. 
The results for both qutrits and ququints show excellent overlap between the experiment and the expected performance---red solid line---of self-guided tomography within the uncertainty margins.
The straight line asymptotic behaviour after the ${\sim} 10^{\text{th}}$ iteration highlights the relationship is a power law, as expected. 
This means that we can get arbitrarily close to the true state at the expense of more iterations. 
Note that we achieve fidelities of $99.92~\substack{+0.04 \\ -0.17}\,\%$ for the qutrit (after 100 iterations), and $99.92~\substack{+0.04 \\ -0.07}\,\%$ for ququint (after 200 iterations), respectively.

The quvigint $(\text{d}{=}20)$ outcomes are similar:
Fig.~\ref{fig:SGQT_LN}c shows the  experimental result (blue dashed line) of self-guided tomography for 20 randomly chosen states of a quvigint compared with simulations (red solid line) for 1000 random states. 
We achieve a fidelity of $99.1\substack{+0.2 \\ -0.6}\,\%$ after 600 iterations. 
The slight deviation of the experimental results from the theory is attributed to the reduced capability of the model to capture all noise sources with increasing dimensions: crosstalk dominates mainly between modes adjacent to the diagonal while in the simulations the crosstalk is evenly distributed among all modes. 
S5~\cite{Supp} shows more details on the prepare-measure correlations matrices, and a comparison between experiment and simulation. 

\begin{figure*}[t]
\includegraphics[width=\textwidth]{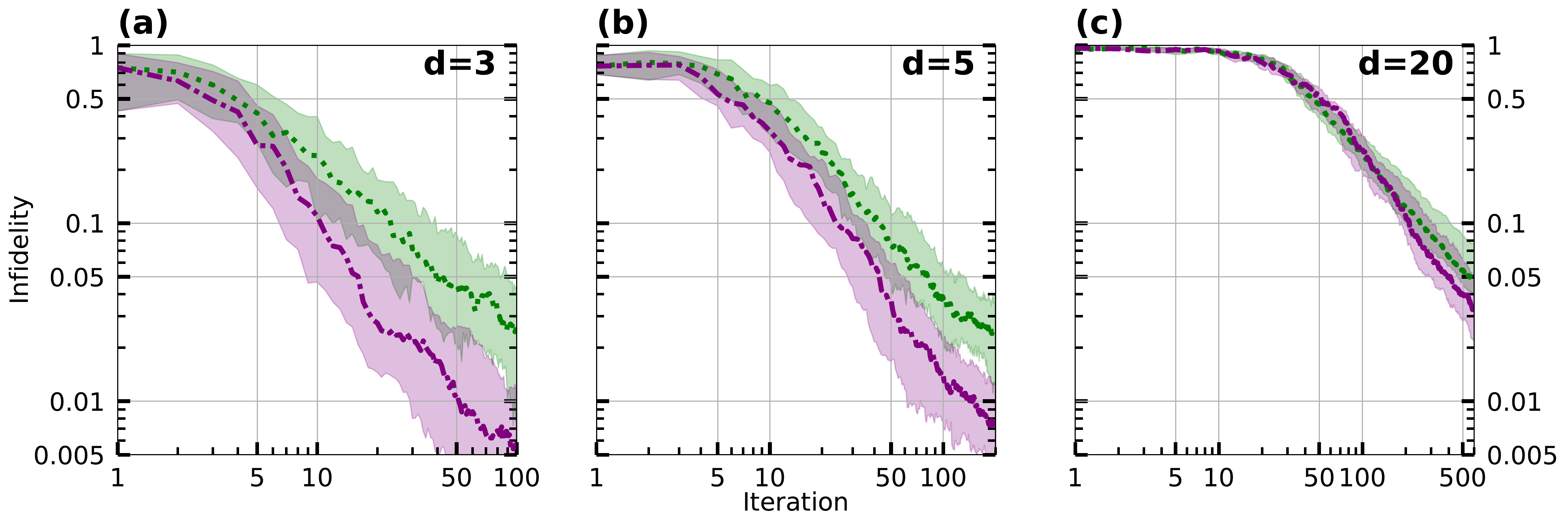}
\vspace{-6mm}
\caption{
\label{fig:SGQT_HNT} 
Self-guided tomography for high statistical noise (dotted green line) and weak atmospheric turbulence (dash-dotted purple line) for (a) qutrit, (b) ququint, and (c) quvigint.
The lines represent the median performance, while the shaded regions are bound by the upper and lower quartile of the infidelities ($(50 \pm 25)\%$). 
The scaling of the horizontal axis is chosen for easy comparison to Fig.~\ref{fig:SGQT_LN}. 
The monotonically decreasing infidelity for statistical and atmospheric noise emphasises the robustness of the algorithm and shows that the true state can be approximated with arbitrary precision using more iterations.
\vspace{-5mm}}
\end{figure*}

We next show the robustness of self-guided tomography in the presence of high statistical noise, which we introduce by reducing the number of state copies $N$ per iteration. 
The self-guided algorithm still performs well in the experiment, reaching fidelities of $98.6\substack{+0.8 \\ -1.3}\,\%$ for qutrits ($N {\sim} 80$, $\Delta N {=} \sqrt{N} {\sim} 9$, after 100 iterations, Fig.~\ref{fig:SGQT_HNT}a), $97.6\substack{+1.2 \\ -1.1}\,\%$ for ququints ($N {\sim} 80$, $\Delta N {\sim} 9$, after 200 iterations, Fig.~\ref{fig:SGQT_HNT}b), and $95.1\substack{+1.0 \\ -2.7}\,\%$ for quvigints ($N {\sim} 1000$, $\Delta N {\sim} 32$, after 600 iterations, Fig.~\ref{fig:SGQT_HNT}c). 
Our results for the qutrit and ququint are consistent with our simulations. 
The simulation for the quvigint faces the same limitations as we discussed above, the comparisons are shown in S4~\cite{Supp}. 

Qudits based on the shape of light are naturally vulnerable to atmospheric turbulence, hence as a third test for self-guided tomography we simulated the effects of free space transmission through the atmosphere. 
Weak turbulence---theoretically approximated as Kolmogorov thin-phase aberrations---has been thoroughly investigated in photons carrying orbital angular momentum~\cite{Paterson2005,Tyler2009,Malik2012}.
These studies agree that turbulence has a strong detrimental effect on the mode profiles, quantified for example by the declining channel capacity with increasing turbulence~\cite{Malik2012}.
The effects are however highly dependent on a set of parameters, including the size of the beam and the scale over which the refractive index correlations remain correlated~\cite{Fried1965}.  
The latter is quantified by the Fried parameter $r_0$ or more commonly the atmospheric turbulence strength $\text{C}_\text{n}^2$. 

Since we are simulating weak turbulence on our SLMs, we ensured we used realistic parameters. 
We note a recent experiment which sent light over 3\,km of strong turbulence in a city~\cite{Krenn2014} that showed only low noise on the phase structure of the beam but significant random displacement of the centre of intensity. 
This beam movement was measured at a kHz rate, hence the average displacement over an integration time of one second remains very small.
Based on this we calculated a phase hologram for an atmospheric turbulence strength of $\text{C}_\text{n}^2 \sim 10^{-18}\,\mathrm{m}^{-2/3}$, corresponding to weak turbulence~\cite{Gbur2008}. 
We chose a transmission distance of ${\sim}1$\,km, resulting in phase value extrema of ${\sim} \pm \pi/5$. 
The selected parameters lead to mode intensity profiles similar to Ref.~\cite{Krenn2014}.
The hologram that simulates turbulence is then displayed on the mSLM along with the hologram for the state.
We generate a new turbulence hologram for each measurement to accurately capture the actual movement of the atmosphere.  
(Additional details and images on the generated holograms are shown in S6~\cite{Supp}.) 

The results of self-guided tomography in the presence of this weak atmospheric turbulence are shown as purple dash-dotted lines in Fig.~\ref{fig:SGQT_HNT}. 
Again, the algorithm shows robustness, leading to fidelities of $99.4\substack{+0.4 \\ -0.6}\,\%$ for qutrits (after 100 iterations, Fig.~\ref{fig:SGQT_HNT}a), $99.2\substack{+0.3 \\ -0.5}\,\%$ for ququints (after 200 iterations, Fig.~\ref{fig:SGQT_HNT}b), and $96.8\substack{+1.0 \\ -1.8}\,\%$ for quvigints (after 600 iterations, Fig.~\ref{fig:SGQT_HNT}c). 
Compared to the low noise results in Fig.~\ref{fig:SGQT_LN}, the infidelities in Fig.~\ref{fig:SGQT_HNT} are around 15 (4) times higher in the high noise (weak turbulence), consistent with our simulations (more details in S7~\cite{Supp}). 
The monotonically decreasing infidelity, despite the presence of considerable noise, strongly suggests that we can get arbitrarily close to the true state with more iterations. 
The robust performance of the algorithm in the presence of noise is especially relevant as attention to qudits in free space quantum communication increases.

\begin{figure}[t]
\includegraphics[width=\columnwidth]{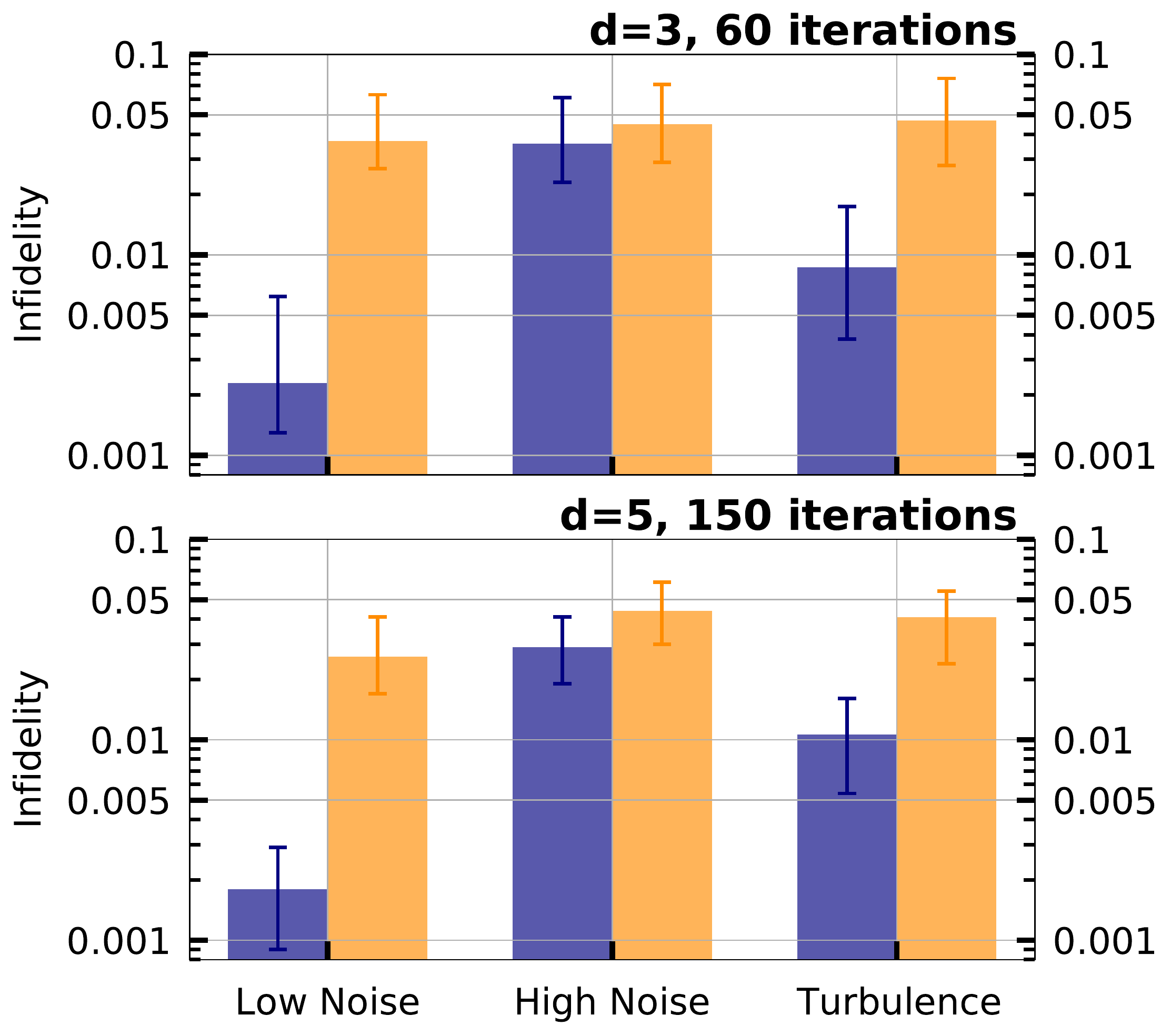}
\vspace{-6mm}
\caption{
\label{fig:VS} 
Comparison of self-guided tomography (blue) with standard tomography (orange) with the same amount of copies of the unknown state. 
Top plot is for a qutrit, bottom plot is for a ququint. 
The left, middle, and right pairs are respectively for low statistical noise, high statistical noise, and weak turbulence. 
Averaging the results for qutrits and ququints, the infidelity of self-guided tomography is lower than standard tomography by a factor of $\approx$ 15, 1.4, and 5, respectively. 
Importantly, the results of self-guided tomography improve with increasing number of iterations---if we iterated for longer the values would be even lower---whereas due to mode-dependent loss in our system standard tomography is fundamentally limited to the values shown in orange. 
\vspace{-6mm}}
\end{figure}

Fig.~\ref{fig:VS} compares qutrit (top row) and ququint (bottom row) self-guided tomography with standard tomography using a fixed amount of copies of the unknown state. 
For the standard tomography, we applied maximum likelihood and root estimators~\cite{Bogdanov2009} to measurements of the full set of mutually unbiased bases (MUBs, e.g. $4~\text{basis sets}~{\times}~3~\text{states} = 12~\text{states}$ for a qutrit~\cite{Giovannini2013}).
It is not practical to perform standard tomography on quvigints, since an over-complete set of measurements as in~\cite{Langford2004, Agnew2011}, would entail implementing 780 measurement holograms and then---unlike self-guided tomography---requires considerable post-processing.

We find that, independent of the dimension, we cannot surpass infidelities in the few-percent range using standard tomography. 
In contrast, self-guided tomography does not display this limitation as already demonstrated in the asymptotic behaviours in Fig.~\ref{fig:SGQT_LN} and Fig.~\ref{fig:SGQT_HNT}. 
Self-guided tomography is also more efficient: for the same amount of resources (copies of the state) it achieves a lower infidelity compared to standard tomography. 
The average improvement factor, as highlighted in Fig.~\ref{fig:VS}, is $\approx$\,15x,\,1.4x and 5x for low noise, high noise, and weak turbulence, respectively.
Equivalently, self-guided tomography needs less resources than standard tomography to reach the same infidelity. 
A comprehensive table comparing the two can be found in S7~\cite{Supp}. 
We have experimentally shown a regime where self-guided tomography alleviates tomography requirements in a photonic system and our results suggest it can be applied to other systems where preparation fidelities is very high but the capacity to quickly create many copies is limited, e.g. in ions.

To extend the applicability of self-guided tomography---which was originally proposed for pure quantum states---we also show its effectiveness for finding qutrit mixed states. 
The modifications we made to the algorithm and details of the experiment are described in S8~\cite{Supp}. 
We reach an average fidelity of $96.5~\substack{+1.1 \\ -3.2}\,\%$ for the case of low statistical noise.  
To demonstrate robustness, we lowered the photon count rate and achieved fidelities of $94.9~\substack{+2.2 \\ -2.9}\,\%$. 
Analogue to the case of pure states, achieving these high fidelities did not require any post-processing. 
It will be worthwhile to further investigate the performance of self-guided tomography on mixed states in the presence of even higher statistical noise or other noise sources, as well as for dimensions beyond the qutrit.

We have shown the advantages of using self-guided tomography for pure states up to 20 dimensions and qutrit mixed states. 
The overall behaviour of self-guided tomography is monotonically decreasing for all tested dimensions, see Fig.~\ref{fig:SGQT_LN} and Fig.~\ref{fig:SGQT_HNT}, indicating the possibility of arbitrarily close state estimation for increasing number of iterations.
This high robustness to experimental and environmental imperfections as well as statistical noise establishes self-guided tomography as a promising candidate for more complex systems, be it multiple qubits or qudits, implemented on any quantum mechanical system.
The same cannot be said of standard tomography especially when applied to systems that are susceptible to mode-dependent losses~\cite{Qassim2014}. 
For example, Ref.~\cite{Bouchard2018}, also using photonic shape qudits, reports a fidelity of $95.3\%$ for a ququint and $93.8\%$ for a 19-level qudit.
We have shown that self-guided tomography is the prime candidate for state tomography in systems of high purity. 

Self-guided tomography also works for processes---a qubit process has been experimentally characterised using this approach~\cite{Hou2020}---and an obvious extension is to $SU(d)$ processes. 
We highlight that self-guided tomography is agnostic to the underlying physical architecture, and hence is applicable to any high-dimensional quantum system, be it spin $N/2$ or phase qudits or an array of entangled qubits. 

\noindent \textit{Acknowledgements}. We thank Kaumudibikash Goswami for helpful discussions on the simulations of atmospheric turbulence. 
This research was supported by the Australian Research Council Centre of Excellence for Engineered Quantum Systems (EQUS, CE170100009). 
JR is supported by a Westpac Bicentennial Foundation Research Fellowship. 
CF acknowledges funding from the Australian Research Council Discovery Early Career Researcher Award (No. DE160100821).


%

\end{document}